\documentclass[a4paper,12pt]{article}

\usepackage[utf8]{inputenc} 
\usepackage[english]{babel} 
\usepackage{graphicx}       
\usepackage{amsmath}        
\usepackage{hyperref}       
\usepackage{geometry}       
\usepackage{setspace}       
\usepackage{cite}  
\usepackage{subcaption} 
\usepackage{amssymb}

\title{(De)-Indexing and the Right to be Forgotten}

\author{
    Salvatore Vilella\textsuperscript{1}, Giancarlo Ruffo\textsuperscript{1}\thanks{Corresponding author: giancarlo.ruffo@uniupo.it} \\
    \textsuperscript{1}\small DISIT, Università del Piemonte Orientale
}

\date{\today}

\begin{document}

\maketitle

\begin{abstract}
In the digital age, the challenge of forgetfulness has emerged as a significant concern, particularly regarding the management of personal data and its accessibility online. The right to be forgotten (RTBF) allows individuals to request the removal of outdated or harmful information from public access, yet implementing this right poses substantial technical difficulties for search engines. This paper aims to introduce non-experts to the foundational concepts of information retrieval (IR) and de-indexing, which are critical for understanding how search engines can effectively "forget" certain content. We will explore various IR models, including boolean, probabilistic, vector space, and embedding-based approaches, as well as the role of Large Language Models (LLMs) in enhancing data processing capabilities. By providing this overview, we seek to highlight the complexities involved in balancing individual privacy rights with the operational challenges faced by search engines in managing information visibility.
\end{abstract}

\tableofcontents
\newpage

\section{Introduction}

De-indexing is the process of removing specific content or links from a search engine's index, preventing them from appearing in search results. Unlike deleting content from a Website, de-indexing does not erase the information from its original source; instead, it ensures that the information is less accessible through search engines. This process is often requested by individuals who seek to limit the visibility of certain online information that may be outdated, irrelevant, or harmful to their reputation.

De-indexing is closely associated with the right to be forgotten (RTBF), a legal principle that allows individuals to request the removal of personal data from public access when it no longer serves a legitimate purpose. Originating in Europe with the landmark case Google Spain SL, Google Inc. v Agencia Española de Protección de Datos in 2014, the RTBF is now a cornerstone of privacy rights in the digital age, defined in legislations such as the General Data Protection Regulation (GDPR), Article 17. De-indexing serves as a practical mechanism for enforcing the RTBF, balancing the individual's right to privacy and the public's right to information.

However, to fully understand the dynamics, the possibilities and the technical challenges behind Web de-indexing, it is important to have a grasp of the basic principles of indexing and, more in general, of information retrieval (IR). The goal of the present document is to provide the reader with a general, technical overview of IR, enabling reflections and allowing an analysis of the implications of de-indexing and RTBF on search engine algorithms and data privacy.

The present document is organized as follows. In Sect.~\ref{sec:models} we provide a technical overview of the most common models of information retrieval, reviewing the basics of boolean models, probabilistic models, vector space and embedding-based models. We will also introduce the reader to Large Language Models (LLMs), which are the latest innovation in IR and textual data mining. Finally, in Sect.~\ref{sec:deindex} we conclude with a brief discussion on deindexing in light of the technical arguments elaborated in the previous sections.

If the reader is particularly interested in IR, we strongly suggest the following references: \cite{manning2008introduction, buttcher2016information, baeza1999modern, witten1999managing}. Nevertheless, this paper succinctly addresses some of the most fundamental aspects behind the principal IR technologies, although our primary objective is to describe how modern IR engines deal with 'forgetting' information.

\section{Models of Information Retrieval}\label{sec:models}

Information retrieval (IR) is the process of finding relevant information from large collections of unstructured or semi-structured data in response to user queries. Unlike structured data retrieval in databases, IR primarily deals with text-heavy datasets such as documents, Web pages, and multimedia. The goal of an IR system is to satisfy the user's information needs by providing results that are relevant and ranked in importance order. This field lies at the intersection of computer science, linguistics, and cognitive science, drawing on various methods to develop algorithms capable of processing and interpreting human language, an inherently complex task.

The fundamental process of information retrieval involves several stages, including preprocessing the data (tokenization, stemming, stop-word removal), indexing the content to enable fast searches, and ranking documents based on their relevance to a query. Modern IR systems utilize a variety of techniques, such as query expansion, relevance feedback, and term weighting, to improve the quality of search results. Increasingly, machine learning (ML) and natural language processing (NLP) are being integrated into IR pipelines, allowing systems to handle user queries more intelligently and adapt to evolving information needs.  

IR models, which provide the theoretical foundation for how information is retrieved, can be broadly divided into three categories:  

\begin{enumerate}
\item Boolean Models: These models treat documents and queries as sets of terms and use strict matching criteria based on logical operators (AND, OR, NOT) to determine relevance. Although simple and intuitive, Boolean models lack mechanisms for ranking results or accounting for partial matches.  
\item Vector Space Models: These models represent documents and queries as vectors in a multi-dimensional space, where each dimension corresponds to a term. Relevance is calculated using measures such as cosine similarity, which captures the degree of alignment between document and query vectors. This approach enables partial matching and ranking, but requires careful term weighting to be effective.  
\item Probabilistic Models: Probabilistic models predict the likelihood that a document is relevant to a given query, based on prior probabilities and the presence or absence of terms. One prominent example is the probabilistic relevance model, which underpins advanced techniques like BM25~\cite{robertson2009probabilistic}.  
\end{enumerate}

In addition to these foundational models, Large, pre-trained, transformer-based deep language models such
as BERT~\cite{devlin2018bert}, T5~\cite{raffel2020exploring} and GPT~\cite{radford2019language}, have been shown effective for text passage retrieval and ranking~\cite{cicero2020beyond,lin2020pretrained,nogueira2019passage,zhuang2021deep}.

\subsection{Boolean Models and Document Representations}

The Boolean retrieval model is a model for information retrieval in which we can pose any query in the form of a Boolean expression of terms, i.e., in which terms are combined with the operators AND, OR, and NOT. In Boolean models, documents and queries are represented as sets of terms and, by applying one or a combination of the logical operators, the model yields the documents that are pertinent to the terms specified in the query. 

There are many possible ways to explore a document: the most intuitive would be a linear scanning (\textit{grepping}) of the entire document to retrieve the selected words. However, this is an extremely expensive process, particularly for very long texts and complex queries. As a response to this issue, we can resort to different techniques. A possible solution would be to build a \textbf{term-document matrix} (TDM). A TDM is a tabular representation used to capture the frequency or presence of terms across a collection of documents. In this matrix each row represents a unique term from the corpus, while each column corresponds to a document. The cell values indicate the frequency (or binary presence) of a term in a document.

Let:
\begin{itemize}
    \item  $\mathcal{C}$ be a corpus consisting of n documents:$D_1, D_2, \ldots, D_n$.
    \item  $\mathcal{V}$ be the vocabulary of unique terms in the corpus, consisting of m terms: $t_1, t_2, \ldots, t_m$.

\end{itemize}

The TDM is an $m \times n$ matrix $\mathbf{A} = [a_{ij}]$, where:
\[
a_{ij} =
\begin{cases} 
    f(t_i, D_j) & \text{if term } t_i \text{ occurs in document } D_j, \\
    0 & \text{otherwise.}
\end{cases}
\]

Here, $f(t_i, D_j)$ is a function that quantifies the association between term $t_i$ and document $D_j$. Common choices for $f(t_i, D_j)$ include:
\begin{enumerate}
    \item \textbf{Term Frequency (TF)}: $f(t_i, D_j) = \text{Count of } t_i \text{ in } D_j$.
    \item \textbf{Binary Representation}: $f(t_i, D_j) = 1 \text{ if } t_i \text{ appears in } D_j; 0 \text{ otherwise.}$
    \item \textbf{TF-IDF (Term Frequency-Inverse Document Frequency)}: \\ $f(t_i, D_j) = \text{TF-IDF score of } t_i \text{ in } D_j$.
\end{enumerate}

As an example, consider a corpus with $\mathcal{C} = \{D_1, D_2, D_3\}$ and vocabulary $\mathcal{V} = \{\text{"apple"}, \text{"banana"}, \text{"orange"}\}$:
\begin{itemize}
    \item $D_1 = \text{"apple banana apple"}$
    \item $D_2 = \text{"banana orange"}$
    \item $D_3 = \text{"orange apple orange"}$
\end{itemize}

The term-document matrix $\mathbf{A}$ with term frequencies is:
\[
\mathbf{A} =
\begin{bmatrix}
    2 & 0 & 1 \\
    1 & 1 & 0 \\
    0 & 1 & 2
\end{bmatrix}
\]

By transforming a corpus of documents into a matrix, algebraic techniques can be leveraged to analyze them more efficiently; moreover, in a real-world corpus the matrix will likely be sparse, enhancing even more computational efficiency.

Another very common method to index documents is the \textbf{inverted index} that provides significant advantages over TDM in terms of space efficiency, query processing speed, support for complex queries, ease of updates, and scalability, making it the preferred data structure choice for modern information retrieval systems.

The inverted index maps each term $t_i \in \mathcal{V}$ to a set of document identifiers and, optionally, the positions where the term appears. Formally, the inverted index is defined as:
\[
\text{Index}(t_i) = \{ (j, \mathcal{P}_{ij}) \mid t_i \text{ occurs in } D_j \},
\]
where $j$ is the index of a document $D_j$ in $\mathcal{C}$ and $\mathcal{P}_{ij}$ is the set of positions where the term $t_i$ appears in document $D_j$.

If we consider the same corpus $\mathcal{C}$ and vocabulary $\mathcal{V}$ of the previous example:
\begin{itemize}
    \item $D_1 = \text{"apple banana apple"}$
    \item $D_2 = \text{"banana orange"}$
    \item $D_3 = \text{"orange apple orange"}$
\end{itemize}

The inverted index for this corpus is:

\[
\text{Index}(\text{"apple"}) = \{(1, \{1, 3\}), (3, \{2\})\}
\]
\[
\text{Index}(\text{"banana"}) = \{(1, \{2\}), (2, \{1\})\}
\]
\[
\text{Index}(\text{"orange"}) = \{(2, \{2\}), (3, \{1, 3\})\}
\]

Here:
\begin{itemize}
    \item $\text{"apple"}$ occurs in $D_1$ at positions $\{1, 3\}$ and in $D_3$ at position $\{2\}$.
    \item $\text{"banana"}$ occurs in $D_1$ at position $\{2\}$ and in $D_2$ at position $\{1\}$.
    \item $\text{"orange"}$ occurs in $D_2$ at position $\{2\}$ and in $D_3$ at positions $\{1, 3\}$.
\end{itemize}

As we can see, while the TDM directly associates terms with numerical values in each document, the inverted index reverses the structure, associating terms with the documents (and positions) they occur in. This structure is also particularly useful for efficient search queries, and is an extremely common method for Web pages indexing.

\subsubsection{Limitations of Boolean Query Models }

Boolean query models, while foundational in traditional IR systems, have several notable limitations:

\begin{itemize}
    \item \textbf{Rigidity of Queries:} Boolean queries are often too rigid as they strictly match the specified keywords without considering their relevance or context within documents. As a result:
    \begin{itemize}
        \item Documents containing all keywords are retrieved, regardless of the relationships or importance of these terms in the document.
        \item Relevant documents that use synonyms or paraphrased expressions might be missed.
    \end{itemize}

    \item \textbf{Feast or Famine Problem:} This issue arises when Boolean queries retrieve either too many documents (feast) or none at all (famine). Specifically:
    \begin{itemize}
        \item Broad queries result in an overwhelming number of results with little relevance.
        \item Narrow queries exclude potentially relevant documents.
    \end{itemize}
    Incorporating proximity-based evaluation in inverted index structures can partially address this issue by ranking documents based on the closeness of query terms.

    \item \textbf{Lack of Flexibility:} Boolean models do not account for semantic similarity or linguistic variations, such as:
    \begin{itemize}
        \item \textbf{Synonyms:} For example, "car" and "automobile."
        \item \textbf{Morphological Variations:} For example, "run" and "running."
    \end{itemize}
    Traditionally, this is addressed by expanding the query vocabulary using techniques such as thesauri or stemming. However, these methods often fail to fully capture nuanced relationships.

    \item \textbf{Binary Results:} Boolean models categorize documents as either:
    \begin{itemize}
        \item Relevant (if they match the query precisely), or
        \item Non-relevant (if they fail to match any part of the query).
    \end{itemize}
    This binary classification does not allow for ranking documents based on their relevance. Advanced IR systems often use ranking mechanisms like Term Frequency-Inverse Document Frequency (TF-IDF) or machine learning models to overcome this limitation.
\end{itemize}

\subsection{Vector Space Models}

The \textbf{Vector Space Model (VSM)} is another possible approach for representing and comparing text data. In this model, both documents and queries are represented as vectors in a high-dimensional space, where each dimension corresponds to a unique term in the vocabulary. The relevance of a document to a query is determined by the similarity between their respective vectors.

Let:
\begin{itemize}
    \item $\mathcal{C} = \{D_1, D_2, \ldots, D_n\}$ be a collection of $n$ documents.
    \item $\mathcal{V} = \{t_1, t_2, \ldots, t_m\}$ be the vocabulary of unique terms in the corpus.
\end{itemize}

Each document $D_j$ is represented as a vector:
\[
\mathbf{d}_j = [w_{1j}, w_{2j}, \ldots, w_{mj}],
\]
where $w_{ij}$ is the weight of term $t_i$ in document $D_j$. The term weights, $w_{ij}$, play a crucial role in determining the effectiveness of the VSM. Common approaches include:
\begin{enumerate}
    \item \textbf{Term Frequency (TF):} Reflects how often a term appears in a document.
    \[
    w_{ij} = \text{TF}(t_i, D_j) = \text{Count of } t_i \text{ in } D_j.
    \]
    \item \textbf{TF-IDF (Term Frequency-Inverse Document Frequency):} Balances term frequency with the rarity of a term across the corpus.
    \[
    w_{ij} = \text{TF}(t_i, D_j) \times \log\left(\frac{n}{\text{DF}(t_i)}\right),
    \]
    where $\text{DF}(t_i)$ is the document frequency of $t_i$ (i.e., the number of documents that contain $t_i$), and $n$ is the total number of documents in the corpus.
\end{enumerate}

Finally, the relevance of a document to a query is computed using similarity measures. The most widely used is \textbf{cosine similarity}, defined as:
\[
\text{CosineSimilarity}(\mathbf{d}_j, \mathbf{q}) = \frac{\mathbf{d}_j \cdot \mathbf{q}}{\|\mathbf{d}_j\| \|\mathbf{q}\|},
\]
where:
\begin{itemize}
    \item $\mathbf{d}_j \cdot \mathbf{q}$ is the dot product of the vectors,
    \item $\|\mathbf{d}_j\|$ and $\|\mathbf{q}\|$ are their magnitudes.
\end{itemize}

Intuitively, cosine similarity measures the angle between the vectors in a $|\mathcal{V}|$-dimensional space, with smaller angles indicating higher similarity between the document and the query.



\subsection{Probabilistic Models}

Finally, let's discuss probabilistic models. Probabilistic models for information retrieval provide a statistical framework to predict the likelihood that a document is relevant to a given query. These models are grounded in probability theory and use the principle of uncertainty to handle the variability and ambiguity inherent in natural language. The fundamental idea is to \textit{rank} documents based on their \textit{probability of relevance} to a query.

\subsubsection{The Probabilistic Relevance Model}

The probabilistic relevance model (PRM) assumes that there exists an ideal set of relevant documents for a query. The goal is to rank documents in order of their probability of belonging to this set. Given a query $Q$ and a document $D_j$, the relevance score is calculated as the probability $P(R=1 \mid D_j, Q)$, where $R$ is a binary variable indicating relevance.

Using Bayes' theorem:
\[
P(R=1 \mid D_j, Q) = \frac{P(Q \mid D_j, R=1) \cdot P(R=1 \mid D_j)}{P(Q)}.
\]

Since $P(Q)$ is constant for a given query, it can be omitted when ranking documents. This leads to the ranking function:
\[
\text{Score}(D_j) \propto P(Q \mid D_j, R=1) \cdot P(R=1 \mid D_j).
\]

To estimate $P(Q \mid D_j, R=1)$, PRM typically assumes independence between query terms:
\[
P(Q \mid D_j, R=1) = \prod_{t \in Q} P(t \mid D_j, R=1).
\]

For practical implementations, simplifying assumptions are made regarding the distribution of terms in relevant and non-relevant documents.

\subsubsection{BM25 and Term Weighting}

As an example of a probabilistic model, let's analyse the BM25 algorithm~\cite{robertson2009probabilistic}. Building on the probabilistic relevance model, BM25 introduces a ranking function widely used in modern IR systems. BM25 addresses two important aspects:
\begin{enumerate}
    \item \textbf{Term Frequency Saturation:} Unlike TF-IDF, where term frequency increases linearly, BM25 applies a saturation function to limit the impact of very frequent terms.
    \item \textbf{Document Length Normalization:} Longer documents are penalized to prevent artificially high scores due to term occurrences.
\end{enumerate}

The BM25 score for a document $D_j$ and query $Q$ is given by:
\[
\text{BM25}(D_j, Q) = \sum_{t \in Q} \text{IDF}(t) \cdot \frac{\text{TF}(t, D_j) \cdot (k_1 + 1)}{\text{TF}(t, D_j) + k_1 \cdot \left(1 - b + b \cdot \frac{\text{len}(D_j)}{\text{avglen}}\right)},
\]
where:
\begin{itemize}
    \item $\text{IDF}(t)$ is the inverse document frequency of term $t$.
    \item $\text{TF}(t, D_j)$ is the term frequency of $t$ in document $D_j$.
    \item $k_1$ and $b$ are tuning parameters controlling term frequency saturation and document length normalization.
    \item $\text{len}(D_j)$ is the length of document $D_j$, and $\text{avglen}$ is the average document length in the collection.
\end{itemize}

\subsection{Document and Word Embeddings}

While traditional VSM represent documents and queries as sparse vectors based on raw term frequencies or TF-IDF weights, this approach often fails to capture semantic relationships between terms. To address this limitation, modern information retrieval systems increasingly rely on \textbf{embeddings}, dense vector representations that encode semantic meaning in a continuous, low-dimensional space.

\subsubsection{Word Embeddings}
\textbf{Word embeddings} are dense vector representations of words, where semantically similar words are mapped to nearby points in the embedding space. These embeddings are typically learned using neural network-based models trained on large corpora of text. Popular methods for generating word embeddings include:

\begin{enumerate}
    \item \textbf{Word2Vec}~\cite{mikolov2013efficient}: Uses a shallow neural network to generate embeddings via two main approaches:
    \begin{itemize}
        \item \textit{Continuous Bag of Words (CBOW):} Predicts a word given its surrounding context.
        \item \textit{Skip-gram:} Predicts the context words given a target word.
    \end{itemize}
    \item \textbf{GloVe (Global Vectors for Word Representation)}~\cite{pennington2014glove}: Combines global word co-occurrence statistics with local context to produce embeddings.
    \item \textbf{FastText}~\cite{bojanowski2017enriching}: Extends Word2Vec by representing words as a collection of character n-grams, allowing better handling of out-of-vocabulary words and morphological variations.
\end{enumerate}

In these models, each word $w_i$ is represented as a vector $\mathbf{v}_i$ in a continuous space, such that the cosine similarity between two vectors approximates the semantic similarity of the corresponding words. For example:
\[
\text{CosineSimilarity}(\mathbf{v}_{\text{king}}, \mathbf{v}_{\text{queen}}) > \text{CosineSimilarity}(\mathbf{v}_{\text{king}}, \mathbf{v}_{\text{car}}).
\]

\subsubsection{Document Embeddings}
While word embeddings capture semantic relationships at the word level, \textbf{document embeddings} provide vector representations for entire documents. These embeddings aggregate information from individual word embeddings or sentences to capture the overall semantic content of a document. Techniques for generating document embeddings include:

\begin{enumerate}
    \item \textbf{Averaging Word Embeddings:} A simple approach where a document embedding is computed as the mean of its constituent word embeddings. While computationally efficient, this method may fail to capture complex relationships and context.
    \item \textbf{Doc2Vec}~\cite{le2014distributed}: Extends Word2Vec by introducing document-specific vectors. Two main approaches are used:
    \begin{itemize}
        \item \textit{Distributed Memory (DM):} Learns document vectors by predicting words based on the document vector and surrounding word vectors.
        \item \textit{Distributed Bag of Words (DBOW):} Learns document vectors by predicting document-specific words without using context.
    \end{itemize}
    \item \textbf{Transformers:} Pretrained language models like BERT~\cite{devlin2018bert}, T5~\cite{raffel2020exploring}, and GPT~\cite{radford2019language} produce powerful contextual embeddings for documents. These models encode documents using self-attention mechanisms, enabling them to capture intricate relationships between words and phrases. For document retrieval, embeddings are typically generated by taking the output of a special token, such as [CLS] in BERT, or by pooling the output embeddings of all tokens.
\end{enumerate}

Overall, document and word embeddings have revolutionized information retrieval by enabling semantic understanding and improving retrieval accuracy. They have been deployed in many tasks, such as:

\begin{itemize}
    \item \textbf{Semantic Search:} Embeddings allow IR systems to retrieve documents that are semantically related to the query, even if they do not share exact terms. For example, a query about "french fries" could retrieve documents about french fries but also about semantically related concepts, such as "cheeseburgers" or other forms of fries; however, it should not suggest other "french" items, that are semantically unrelated to the query.    
    \item \textbf{Ranking and Similarity Scoring:} Dense embeddings enable more accurate scoring of document-query relevance, leveraging metrics such as cosine similarity or neural network-based ranking models.
    \item \textbf{Clustering and Classification:} Embeddings facilitate document clustering and topic modeling, grouping semantically similar documents together.
\end{itemize}

A geometrical representation of the semantic space allows automated systems to have a deeper understanding of the textual context, improving their performances and the user-experience.

\subsection{Large Language Models}\label{sec:llms}

\begin{figure}[htbp]
    \centering
    \begin{minipage}[c]{0.45\textwidth} 
        \centering
        \vspace*{5mm} 
        \includegraphics[width=\linewidth]{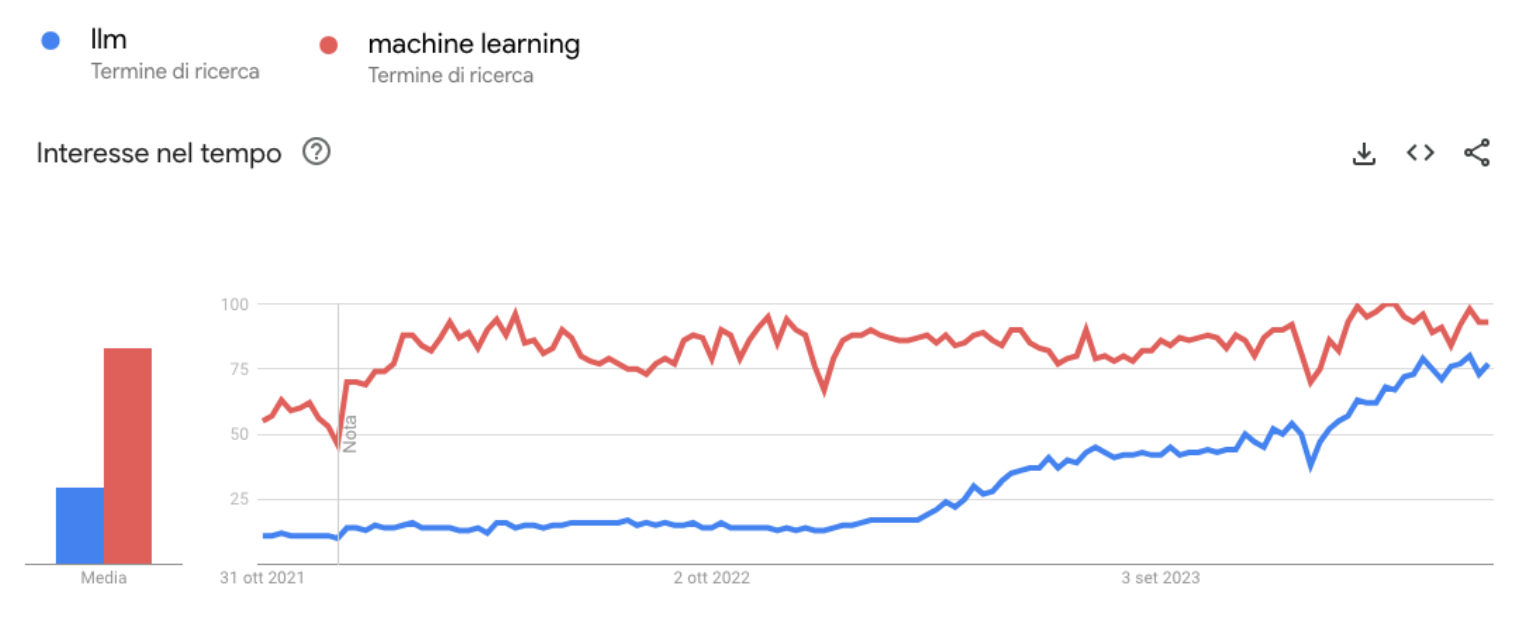}
        \label{fig:interest}
    \end{minipage}
    \hfill
    \begin{minipage}[c]{0.45\textwidth} 
        \centering
        \includegraphics[width=\linewidth]{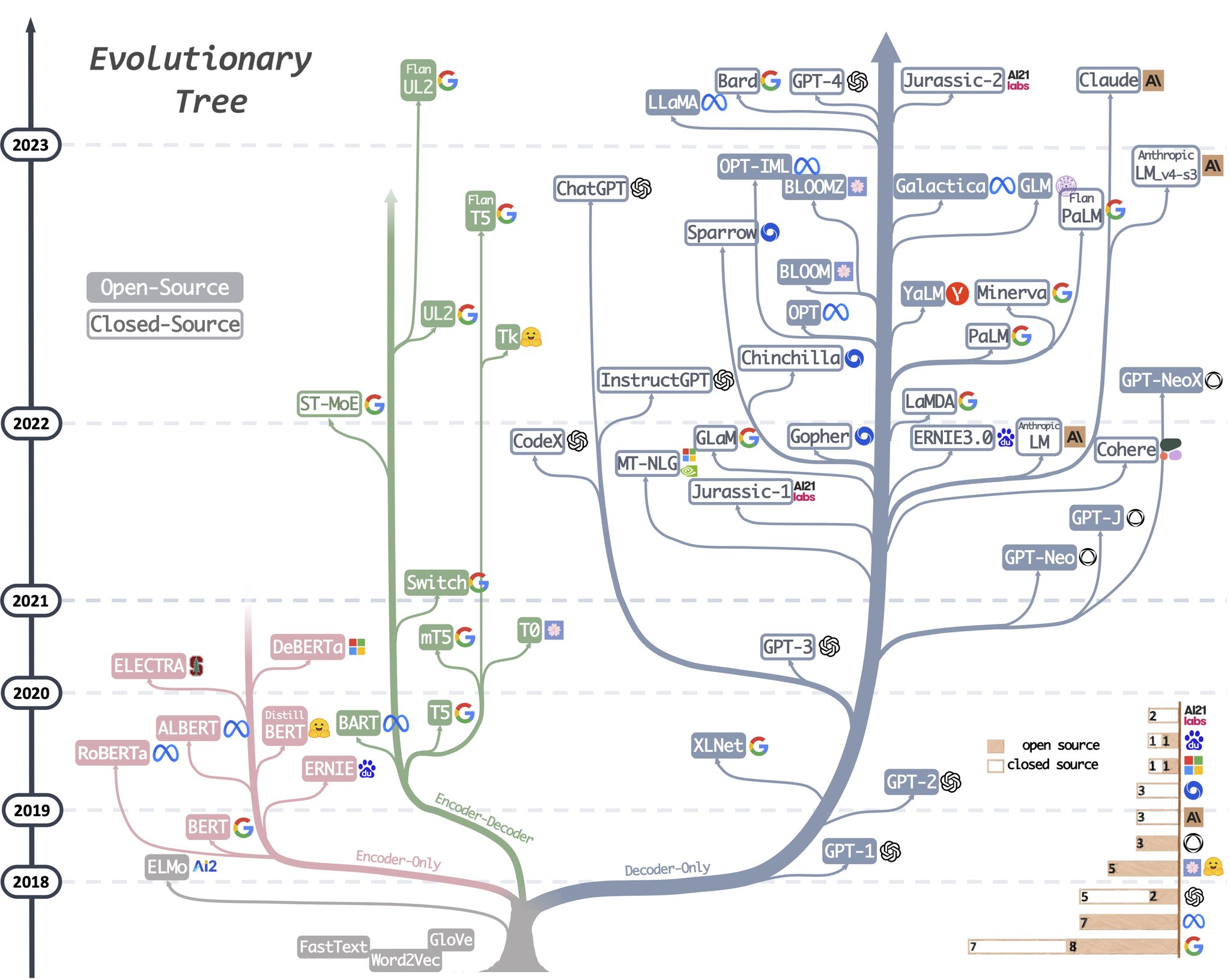}
        \label{fig:evotree}
    \end{minipage}
    \caption{\textbf{Left}: interest in time of the queries \textit{machine learning} and \textit{llm} on Google. We can see how LLMs gain momentum and approach the level of the machine learning query over time. \textbf{Right}: the evolutionary tree of LLMs~\cite{yang2023harnessing}. }
    \label{fig:fig1}
\end{figure}

Large language models (LLMs) are the meeting point of vector space models and deep neural networks. As we can see in Fig.~\ref{fig:fig1}, during the last years they gained a huge momentum, becoming progressively more complex as they get more embedded in our daily lives. They are indeed advanced neural networks designed to process and generate human-like text by learning from large-scale datasets and leveraging sophisticated architectures such as transformers. The central concept behind these models is their ability to represent and manipulate information in a high-dimensional \textbf{parameter space}, denoted mathematically as \(\theta \in \mathbb{R}^d\), where \(d\) is the number of parameters. These parameters, which can number in the hundreds of billions, encode the weights and biases of the neural network, allowing it to model complex patterns in natural language.

\begin{figure}
    \centering
    \includegraphics[width=0.5\linewidth]{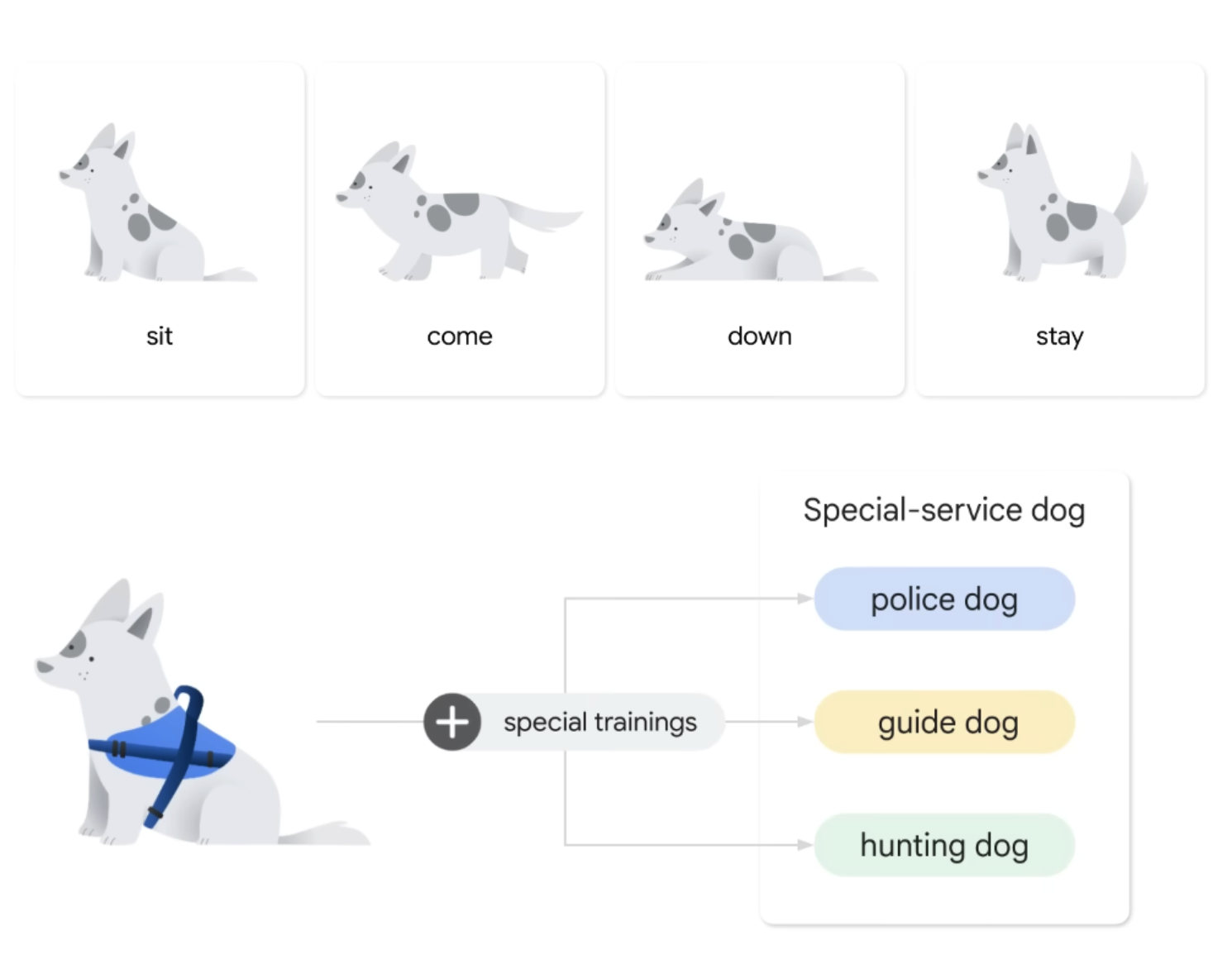}
    \caption{An intuitive comparison between the steps required to train a special-service dog and the training and fine-tuning phases of a LLM~\cite{google_intro_llms}.}
    \label{fig:dogtraining}
\end{figure}

LLMs, just as any other supervised deep learning model, undergo an initial training phase, and a second phase of fine tuning. As Google puts it in its Introduction to LLMs~\cite{google_intro_llms}, these phases can be intuitively visualised as the steps required to train a special-service dog, as in Fig.~\ref{fig:dogtraining}. At a first stage, it will be necessary to teach the dogs the basics such as sitting, staying, or responding to commands—skills that are broadly applicable and form the foundation of their training. This is analogous to the initial training phase of LLMs, where the model learns general patterns and structures of language from vast datasets.  

In the second phase, fine-tuning, the focus shifts to teaching the dog specific tasks, such as guiding a visually impaired person or detecting medical conditions. Similarly, fine-tuning an LLM involves specializing the model for particular applications, such as answering domain-specific questions or generating content tailored to a specific audience.

\subsubsection{Training Phase}

The training phase optimizes the model parameters \(\theta\) to minimize a loss function \(L(\theta)\), typically defined over a large corpus of text. For example, in autoregressive models such as GPT, the objective is to maximize the probability of the next token \(t_i\) given the previous tokens \(t_1, t_2, \ldots, t_{i-1}\). This is expressed as:  
\[
\mathcal{L}(\theta) = -\sum_{i=1}^N \log P_\theta(t_i | t_1, t_2, \ldots, t_{i-1}),
\]  
where \(N\) is the total number of tokens in the dataset. The probability \(P_\theta(t_i | t_1, t_2, \ldots, t_{i-1})\) is typically computed using the transformer architecture~\cite{vaswani2017attention}, which employs multi-head self-attention mechanisms:  
\[
\text{Attention}(Q, K, V) = \text{softmax}\left(\frac{QK^\top}{\sqrt{d_k}}\right)V,
\]  
where \(Q\), \(K\), and \(V\) are query, key, and value matrices derived from input embeddings, and \(d_k\) is the dimensionality of keys. The stacked layers of the transformer allow the model to capture both local and global dependencies throughout the sequence.

\subsubsection{Fine-Tuning Phase}

After pretraining, the model can be fine-tuned for specific tasks. This involves further updating \(\theta\) using a smaller domain-specific dataset \(D_{\text{task}}\). For example, in supervised fine-tuning, the loss function is modified to reflect the task-specific objective:
\[
\mathcal{L}_{\text{task}}(\theta) = -\sum_{(x, y) \in D_{\text{task}}} \log P_\theta(y | x),
\]
where \(x\) and \(y\) are input-output pairs (e.g., questions and answers). Techniques such as Reinforcement Learning with Human Feedback (RLHF)~\cite{christiano2017deep} refine the model by optimizing rewards \(R(\theta)\) derived from human preferences.

\subsubsection{Early and Modern LLMs}

The evolution of LLMs began with models such as OpenAI's GPT~\cite{radford2018improving}, which introduced unsupervised pre-training in large corpora followed by fine-tuning. GPT-2~\cite{radford2019language} scaled this approach with 1.5 billion parameters, while GPT-3~\cite{brown2020language} expanded to 175 billion parameters, allowing for learning with a few shots. These advances rely on efficient parallelization strategies, such as model and data parallelism, to handle the immense computational demands.

Modern models like GPT-4 and Google’s PaLM 2~\cite{anil2023palm} push the boundaries further, incorporating architectural refinements and enhanced training techniques. For example, these models often employ techniques such as dropout regularization, layer normalization, and adaptive learning rate schedules to stabilize training and improve generalization. The size of these models is often measured in scientific notation, with parameter counts reaching \(10^{11}\) or greater.

\section{(De)-Indexing and implications on RTBF}\label{sec:deindex}

Having established a basic technical background on Information Retrieval models, we can now wrap them up and relate them to the issue of content indexing (and de-indexing) from the Web. As we saw, the emergence of neural embeddings has brought significant advancements in the field of information retrieval, particularly in how documents are represented, retrieved, and analyzed. Unlike traditional models such as bag-of-words, which reduce documents to unordered collections of word frequencies and fail to capture contextual or semantic meaning, neural embeddings provide a robust representation of documents within a continuous vector space. This shift enables IR systems to better understand the content and context of documents, unlocking new possibilities for semantic retrieval and analysis.

Document embeddings encode the meaning and relationships of words and phrases, allowing for a deeper and more nuanced representation of textual content. By projecting documents into high-dimensional spaces, embeddings capture semantic similarities that are not easily discernible using keyword-based approaches. This capability is particularly valuable in overcoming the limitations of traditional IR techniques, which often rely on exact keyword matching and struggle with synonyms, paraphrasing, or variations in language. For example, using document embeddings, a query about ``renewable energy sources'' can retrieve documents discussing ``solar power'' or ``wind energy,'' even if the precise terms in the query are absent from the documents.

Another major advantage of document embeddings lies in their ability to compute semantic similarity between documents by leveraging distance metrics, such as cosine similarity, in the embedding space, as we previously described. Such a mechanism enables the retrieval of documents that are semantically related, even when their keywords differ or are arranged in a nonmatching order. For example, two documents discussing the same topic but expressed in unique ways can still be identified as closely related in the embedding space, greatly improving retrieval accuracy and relevance.

Furthermore, neural embeddings exhibit impressive generalization capabilities, particularly when trained on large corpora. These models can effectively apply learned semantic relationships to new, unseen documents, making them highly scalable for dynamic and ever-growing text collections such as those found on the Web. This scalability ensures that IR systems can keep pace with the rapid evolution of content, providing consistent performance and broad coverage even as new topics emerge.

However, as it often happens, \textit{there is no such thing as a free lunch}. Controlling and managing the visibility of information on the Web is a complex task. Completely removing a document is often technically unfeasible, requiring interventions such as DNS modifications or low-level routing policy changes, which can be circumvented. A more practical solution involves \textit{de-indexing} specific documents from major search engines. This approach does not physically remove content from the Web; instead, it reduces the visibility of the targeted documents and disrupts their contextual relationships in the embedding space. For instance, let's consider the case of an individual (Mr. John Smith) who decided to exercise his RTBF by asking major search engines providers to de-index the content related to a court case in which he was involved. Specifically, he will request to take down all the documents that are linking him to the matter of the discussion, that we will indicate by the token "X". If documents linking "Smith" and "X" are de-indexed, their semantic association weakens; on the other hand, however, other associations, such as between "Smith" and another token "Y", may grow stronger. This dynamic adjustment of relationships in the embedding space has profound implications for how information is represented and retrieved: we don't know if the new association with "Y" is, according to Mr. Smith, more or less desirable, and it is not easy to predict the consequences that such an operation can have on the embedding space at a larger scale. 
There are already cases that are, apparently, related to cases of removal of information of individuals from LLMs. A recent example is the case of David Mayer, a prompt that causes ChatGPT to output an error message without any further explanation\footnote{https://shorturl.at/nisj8}. The glitch led to various theories among users, including concerns about privacy and the right to be forgotten. Some speculated that David Mayer himself might have requested his information be removed from ChatGPT’s responses, though this was later labeled by Open AI as a misunderstanding of the technical issue at hand\footnote{https://shorturl.at/KREbP}. Even though this case might not be related to the specific issue of knowledge removal, it highlights how software houses should adopt sophisticated methods to approach the issue, because hard-coded, a-posteriori patches could cause communication issues among the user base.

These considerations inevitably impact forgetfulness and RTBF, having implications on individuals and on \textit{collective memory}. There are several studies that explored, from an empirical and mathematical point of view, the dynamics of collective memory and attention. For instance, in~\cite{candia2019universal} the authors differentiate between \textit{communicative} and \textit{cultural} memory, testing a bi-exponential decay function for attention across various cultural domains, implying that the initial collective attention (reflecting communicative memory) decreases rapidly, followed by a slower decline (reflecting cultural memory). An empirical approach to this topic could indeed be useful to quantify the impact that these new forms of information retrieval, such as LLM-based search engines, can have on collective memory and RTBF. While there are clear positive aspects, such as a quick and (hopefully) accurate retrieval and sharing of information, that are dynamically contextualized based on user interactions, there are also negatives. Many of them are common also to other modern forms of information retrieval, such as the over-reliance on the specific technology, or a homogenization of knowledge that can happen if LLMs tend to prioritize popular or widely accepted narratives while sidelining minority perspectives. What's new, instead, is the deep complexity of the dynamics of machine forgetting, that could go both ways: either in an inadvertent forgetting of knowledge (which is partly what happens in the above mentioned homogenization process) or in the impossibility of complete forgetting, due to the deep entanglement of concepts in the embedding space.

\subsubsection*{Forgetfulness as a problem of \textit{machine unlearning}}

As opposed to a-posteriori de-indexing, we can define the problem of \textit{machine unlearning}. Machine unlearning refers to the process of selectively removing the influence of specific training data points from an already trained machine learning model. The goal of machine unlearning is to enable a model to behave as if it had never encountered certain data, thereby addressing privacy concerns and enhancing model adaptability. An interesting Stanford AI Lab report~\cite{liu2024unlearning} summarizes the past, present and future prospects of machine unlearning, describing different approaches that allow for the removal of irrelevant or outdated information, helping models maintain their performance over time. While it is not a trivial task, it becomes even more challenging within the context of deep learning models and LLMs. Indeed it has been shown how they tend to retain knowledge even after attempts at unlearning, complicating efforts to remove harmful or unwanted capabilities instilled during pretraining~\cite{cossu2022continual, casper2023deep}. 

In this context, Retrieval-Augmented Generation (RAG)~\cite{aws_rag} emerges as a potential approach for balancing the benefits of neural embeddings with the need for greater interpretability and control. RAG systems combine the strengths of semantic retrieval and generative models, enabling them to deliver accurate, context-aware responses while mitigating some of the risks associated with opaque embedding spaces. By integrating retrieval and generation into a unified framework, RAG offers a pathway to enhance the precision and relevance of IR systems while maintaining a degree of oversight and adaptability. Given its characteristics, RAG is being explored as a potential solution that allows for simulated forgetting without direct interaction with the model itself~\cite{wang2024machine, hoang2024learn}. This approach addresses some of the significant limitations of traditional LLMs, such as their tendency to produce inaccurate or outdated information, commonly referred to as "hallucinations". By leveraging external data sources, RAG enables models to generate more accurate and contextually relevant outputs. In general, a RAG framework consists of several interconnected components that work together to retrieve relevant information and generate responses:

\begin{itemize}
\item \textbf{User Query Input}: The process begins when a user submits a query. This input is projected into an LLM embedding space to capture the semantic meaning of the query.

\item \textbf{Retrieval System}:
\begin{itemize}
    \item \textbf{Document Retrieval}: The retrieval system scans \textit{external knowledge bases} (e.g., databases or document collections) to fetch relevant text chunks. This can involve traditional keyword-based methods (sparse retrieval) or modern dense retrieval techniques that utilize embeddings for similarity search.
    \item \textbf{Ranking and Filtering}: After retrieving potential documents, the system ranks them based on relevance, typically selecting the top N documents for further processing.

\end{itemize}

\item \textbf{Contextual Embedding Generation}: Each retrieved document is also converted into an embedding to ensure that the generative model can effectively incorporate this information during response generation.
\item \textbf{Fusion Mechanism}: The retrieved documents are fused with the original query through either early or late fusion methods. In early fusion, both the query and documents are fed into the generative model simultaneously; in late fusion, retrieved documents refine the model's output after initial generation~\cite{rackauckas2024ragfusion}.
\item \textbf{Response Generation}: The LLM generates a coherent response based on the augmented input, leveraging both its pre-existing knowledge and the newly retrieved information.

\end{itemize}

A RAG-based framework can facilitate machine unlearning in several ways, for example by updating in real-time the knowledge base without retraining the entire model. When specific data points need to be forgotten, they can be removed from the external knowledge base. This action effectively simulates unlearning by ensuring that future queries do not retrieve or rely on this data. By managing an external knowledge base, RAG supports continuous learning while enabling selective forgetting of harmful or outdated information. RAG’s reliance on an external retrieval system allows for more efficient updates, where only the knowledge base needs to be modified rather than the model itself. 

These technical considerations intersect with critical ethical and governance questions. The power to decide which documents are de-indexed, and thus rendered less visible, lies predominantly with a few dominant actors in the IR ecosystem, even though it is regulated to different extents by central authorities. This concentration of influence could raise concerns about transparency, accountability, and the potential for abuse. Additionally, the training phase of neural embeddings plays a decisive role in shaping the semantic relationships encoded in the embedding space. However, as we described, this phase is inherently opaque, making it challenging to ensure fairness in the resulting IR systems. 

As also many authoritative sources envision~\cite{liu2024rethinking}, it is very likely that, considering its interesting technical challenges and its crucial and multi-faceted implications on society, forgetfulness will be a prominent research topic in computer science and, more broadly, in AI-related research in the upcoming years.

\section*{Funding} 
This research has been funded by the European Union - Next Generation EU, Mission 4 Component 2 - CUP C53D23005810006, project ``Forgetfulness, between rights, duties and technological possibilities'' (PRIN 2022 - PNRR M4C2).

\bibliographystyle{plain}
\bibliography{references}

\end{document}